\documentclass[aps,pre,twocolumn]{revtex4}

\usepackage{graphicx,amssymb,amsmath}
\usepackage{color} 
\usepackage{multirow}

\newcommand{\be}{\begin{equation}}
\newcommand{\ee}{\end{equation}}
\newcommand{\br}{\mathbf{r}}
\newcommand{\brho}{\mbox{\boldmath${\rho}$}}
\newcommand{\bxi}{\mbox{\boldmath${\xi}$}}
\newcommand{\bt}{\mathbf{t}}
\def\eq#1{Eq.~(\ref{#1})}

\usepackage{array}
\newcolumntype{C}[1]{>{\centering\let\newline\\\arraybackslash\hspace{0pt}}m{#1}}
\newcolumntype{L}[1]{>{\raggedright\let\newline\\\arraybackslash\hspace{0pt}}m{#1}}
\newcolumntype{R}[1]{>{\raggedleft\let\newline\\\arraybackslash\hspace{0pt}}m{#1}}

\begin{document}

\title{Slow closure of denaturation bubbles in DNA: twist matters}

\author{Anil Kumar Dasanna}
\author{Nicolas Destainville}
\author{John Palmeri} 
\altaffiliation{Present address: Laboratoire Charles Coulomb, D\'{e}partement Physique Th\'{e}orique, Universit\'{e} Montpellier 2, F-34095 Montpellier, France, EU}
\author{Manoel Manghi}
\email{manghi@irsamc.ups-tlse.fr}

\affiliation{Universit\'e de Toulouse, UPS, Laboratoire de Physique Th\'eorique (IRSAMC), F-31062 Toulouse, France, EU;
CNRS; LPT (IRSAMC); F-31062 Toulouse, France, EU}

\date{\today}

\begin{abstract}
The closure of long equilibrated denaturation bubbles in DNA is studied using Brownian dynamics simulations. A minimal mesoscopic model is used where the double-helix is made of two interacting bead-spring freely rotating strands, with a non-zero torsional modulus in the duplex state, $\kappa_\phi=200$ to $300\,k_{\rm B}T$. For DNAs of lengths $N=40$ to $100$~base-pairs (bps) with a large initial bubble in their middle, long closure times of 0.1 to $100~\mu$s are found. The bubble starts winding from both ends until it reaches a $\approx 10$~bp metastable state, due to the large elastic energy stored in the bubble. The final closure is limited by three competing mechanisms depending on $\kappa_\phi$ and $N$: arms diffusion until their alignment, bubble diffusion along the DNA until one end is reached, or local Kramers process (crossing over a torsional energy barrier). For clamped ends or long DNAs, the closure occurs via this latter temperature activated mechanism, yielding for the first time a good quantitative agreement with experiments.
\end{abstract}
\maketitle

\section{Introduction}

Since the discovery of the double helical DNA structure by Watson and Crick in 1953~\cite{watsoncrick}, many studies have highlighted  the role played by local DNA winding or unwinding in important biological processes, such as DNA replication, transcription or repair~\cite{alberts}. Biophysical experiments using single-molecule techniques~\cite{smith} have shown that applying an external torque to DNA induces the formation of plectoneme or other structural changes~\cite{marko}. Among them, the nucleation of a DNA denaturation bubble, a segment of several consecutive broken base pairs (bp), has been observed~\cite{dean} and theoretically predicted~\cite{benhamPNAS,anshelevich,benhamPRL,benhamPRE,jost} when a superhelical stress is imposed. 

In this paper, we focus on the role played by DNA torsional elasticity and twist dynamics in the spontaneous closure of equilibrated large denaturation bubbles at room temperature. At first sight, once the bubble has been nucleated, for instance \textit{in vivo} by the help of enzymes, it should close almost instantaneously since the temperature is smaller than the denaturation one. However, very large bubble lifetimes, in the $20-100\,\mu$s range for a 30~bps DNA, have been observed in \textit{in vitro} experiments by Altan-Bonnet \textit{et al.}~\cite{altan-bonnet}. These lifetimes are interpreted as closure times of the central bubble made of 18 AT (Adenosine and Thymine) nucleotides, flanked by two GC (Guanine and Cytosine) arms, known to be more stable.

Several models~\cite{fogedby,bar,chakrabarti,vanerp} have studied the bubble breathing, i.e. intermittent and fast opening/closure of small bubbles, by considering an effective dynamics of the base-pairing states without focusing on the chain degrees of freedom. 
For instance, the Peyrard-Bishop model~\cite{peyrard} has been extended to consider twist degrees of freedom~\cite{barbi,cocco,campa}. This model suffers two strong approximations: (i) the helical axis is kept straight, i.e.. both bending and the chain orientational entropy are neglected; and (ii) local breathing bubbles (or ``breathers'') emerge as localized excitations of a non-linear wave equation which comes from a Hamiltonian dynamics with inertia~\cite{yakushevich}. However, since the dynamics of DNA in water is overdamped, these excitations have a lifetime of a few picoseconds~\cite{alexandrov,peyrard2009}. These approaches are thus only valid at short time scale, such that the chain configuration can be considered as frozen, and cannot explain such large lifetimes as considered here. 

Other numerical works focus on the chain and internal dynamics with various levels of coarse-graining, using molecular dynamics~\cite{knotts2007,pablo2009,mazur,sayar2010} or Langevin dynamics simulations~\cite{schatz2000,jeon,mielke2008,dorfman2011}. However they fail to capture the $\mu$s time scale time due to their high level of precision, or they are limited to short 30~bp long DNA~\cite{zeida}. In a recent paper~\cite{dasanna}, we have proposed a simple coarse-grained model where two semi-flexible strands interact and form a planar ``ladder'' in the dsDNA form (without helicity). The coupling between base-pairing and bending elasticity was introduced through a varying persistence length equal to $\ell_{\rm ds}= 150$~bp for dsDNA, and $\ell_{\rm ss}=3$~bp for single-stranded (ss) DNA~\cite{palmeri2007}. Closure times of $0.1~\mu s$ to $4~\mu s$, following a scaling law of $\tau \approx N^{2.4}$, where $N$ is the DNA length, were found, but still much smaller than the experimental closure lifetimes. 

In this paper, we improve this numerical model so that the two strands interwind to form a double helix in the double-stranded (ds) DNA state. The torsional modulus, $\kappa_\phi$, is chosen between 200 and $300~k_{\rm B}T$ in the dsDNA state (corresponding to torsional rigidity around 2.4 to $4.5\times 10^{-19}$ J\,nm~\cite{smith,benhamPNAS,bouchiat}) and taken to vanish in the bubble. We show that twist dynamics plays a key role in the closure of equilibrated large bubbles, which occurs in two steps. First, the large flexible bubble quickly winds from both ends ({\it zipping} regime), thus storing bending and torsional energy in the bubble, which stops when it reaches a size of $\approx10$~bps. The closure of this metastable bubble depends on $\kappa_{\phi}$ and $N$: for low $\kappa_\phi$, an \textit{arms diffusion limited} (ADL) regime is observed, as in previous ladder model, where the closure is controlled by the diffusive alignment of the two ds arms; for large $\kappa_\phi$ and not too large $N$, the bubble diffuses along the DNA and closes as soon as it reaches one DNA end [\textit{bubble diffusion limited} (BDL) closure], with a closure time in $\tau\approx N^{2.3}$ for $40\le N\le 100$. For large $\kappa_\phi$ and $N$ or clamped ends, the closure is \textit{temperature activated} (TA), which now accounts for the experimental observations~\cite{altan-bonnet}.

\section{Model}

\begin{figure}[!t]
\centering
\includegraphics[width=.4\textwidth]{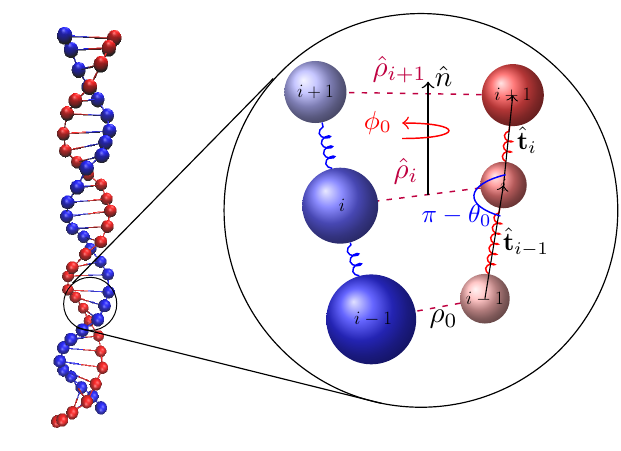}
\caption{\label{fig0} (Color online) Snapshot of an equilibrated double-helix. The bending angle along each strand is $\theta_0$, $\rho_0$ is the equilibrium base-pair distance, and $\hat{n}$ is the helical axis around which twist is defined. The imposed equilibrium twist between successive pairs is $\phi_0$.}
\end{figure}

The DNA is modeled by two interacting bead-spring chains each made of $N$ beads (of radius $0.17$~nm) of position $\br_i$. The Hamiltonian is $\mathcal{H} = \mathcal{H}_{\rm el}^{(1)} + \mathcal{H}_{\rm el}^{(2)} + \mathcal{H}_{\rm tor} + \mathcal{H}_{\rm int}$, where the first two contributions are elastic energies of the strands $j=1,2$ which include both stretching and bending energies 
\be
\mathcal{H}_{\rm el}^{(j)} = \sum_{i=0}^{N-1}\frac{\kappa_{\rm s}}{2}(r_{i,i+1}-a_0)^{2} + \sum_{i=0}^{N-1}\frac{\kappa_{\theta}}{2}(\theta_i-\theta_0)^{2}
\ee
The stretching modulus is $\beta\kappa_{\rm s}=100$,  where $\beta^{-1}=k_{\rm B}T_0$ is the thermal energy, $T_0=300$~K is room temperature, and $a_0=0.357$~nm. The bending modulus is large, $\beta\kappa_{\theta}=600$, to maintain the  angle between two consecutive tangent vectors along each strand, $\theta_i$, to the fixed value $\theta_0=0.41$~rad (see Fig~\ref{fig0}). Each strand is thus modeled as a Freely Rotating Chain (FRC)~\cite{grosberg}.
\begin{figure*}[!t]
\includegraphics[width=\textwidth]{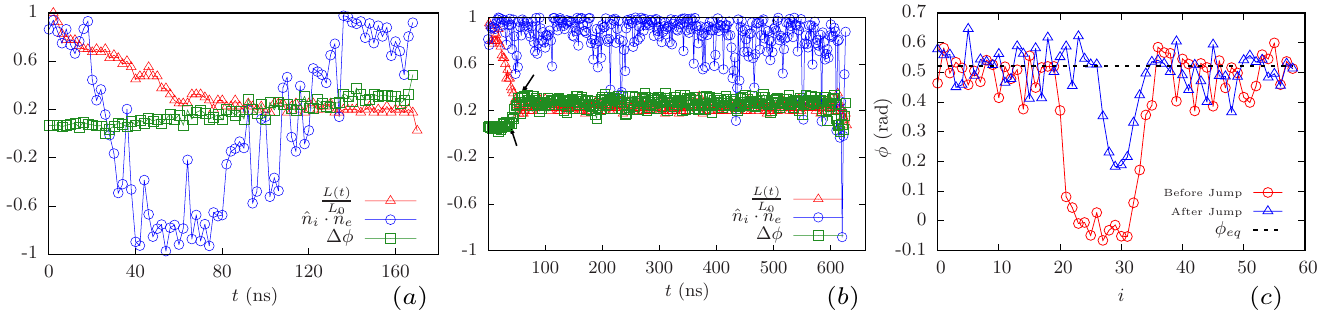}
\caption{\label{fig1} (Color online) Time evolution of the adimensional bubble size $L(t)/L(0)$ (red), the scalar product between the two dsDNA arms axes, ${\bf\hat{n}}_{\rm i}\cdot{\bf\hat{n}}_{\rm e}$, and the average twist angle per bp, $\Delta\phi$, in the bubble for $N=60$ and (a) $\beta \kappa_{\phi}=200$; (b)  $\beta\kappa_{\phi}=300$. (c) Profile of the twist angles in the dsDNA just before ($\circ$) and just after the onset of the metastable regime ($\triangle$), marked by arrows in (b) ($\phi_{\rm eq}=0.52$~rad).}
\end{figure*}
The third and fourth terms of $\mathcal{H}$ are the torsional energy and hydrogen-bonding interactions respectively. The torsional energy is modeled by an harmonic potential, 
\be
\mathcal{H}_{\rm tor} = \sum_{i=1}^{N-1} \frac{\kappa_{\phi,i}}{2}(\phi_i-\phi_0)^2
\ee
where $\phi_i$ is defined as the angle between two consecutive base-pair vectors $\brho_i \equiv \br_{i}^{(1)} - \br_{i}^{(2)}$ and $\brho_{i+1}$  ($\phi_0 = 0.62$~rad).
The stacking interaction between base-pairs is modeled through a $\kappa_{\phi,i}$ that depends on the distances between complementary bases, $\kappa_{\phi,i}= \kappa_{\phi}[1-f(\rho_i)f(\rho_{i+1})]$ where $ f(\rho_i) = [1+\mathrm{erf}\left(\frac{\rho_i-\rho_b}{\lambda'}\right)]/2$, and $\rho_i = | \brho_i |$. Hence, $\kappa_{\phi,i}=\kappa_\phi$ in the dsDNA state, and $\kappa_{\phi,i}=0$ in the ssDNA one. We have chosen $\lambda' = 0.15$~nm and $\rho_b = 1.5$~nm and checked that a slight change in these values does not change significantly the results.
The hydrogen-bonding interaction is modeled by a Morse potential, 
\be
\mathcal{H}_{\rm int} = \sum_{i=1}^{N}A(e^{-2\frac{\rho_i-\rho_0}{\lambda}}-2 e^{-\frac{\rho_i-\rho_0}{\lambda}})
\ee
where $\rho_0=1$~nm, $\lambda=0.2$~nm, and $\beta A=8$ as in Ref.~\cite{dasanna}.

The evolution of $\br_i(t)$ is governed by the over-damped Langevin equation, integrated using an Euler's scheme,
\be
\zeta \frac{d\br_i}{d t} = -\nabla_{\br_i}\mathcal{H}(\{\br_j\})+\bxi_{i}(t) 
\ee
where $\zeta=3\pi\eta a$ is the friction coefficient for each bead of diameter $a$ with $\eta=10^{-3}$~Pa.s the water viscosity. The random force of zero mean, $\bxi_i(t)$, mimics the action of the thermal heat bath and obeys the fluctuation-dissipation relation $\langle \bxi_i(t)\cdot\bxi_j(t') \rangle = 6k_{\rm B}T\zeta\,\delta_{ij}\,\delta(t-t')$. Lengths and energies are made dimensionless in the units of $a=0.34$~nm and $k_{\rm B}T_0$ respectively. The dimensionless time step is $\delta \tau=\delta t k_{\rm B}T_0/(a^2\zeta)$, set to $5\times10^{-4}$ ($\delta t=0.045$~ps) for sufficient accuracy~\cite{dasanna}. The equilibrium properties of this model DNA are described in the Appendix~\ref{AppA}. As an example, a typical equilibrium configuration of a 30~bp dsDNA is shown in Fig.~\ref{fig0}. In particular, the fitted values for the dsDNA persistence length and the pitch are $\ell_{\rm ds}\approx 160$~bp and $p=12$~bp for $\beta \kappa_\phi=300$, which are comparable to the actual dsDNA values ($\ell_{\rm ds}\approx 150$~bp and $p=10.4$~bp). The ssDNA persistence length is $\ell_{\rm ss}=3.7$~nm, compatible with experimental measurements~\cite{tinland} (see Appendix~\ref{AppA}).
The initial bubble, of size $L(t=0)=N-20$, is created in the middle of the DNA by switching off $\mathcal{H}_{\rm int}$ and then equilibrated for $3\,\mu$s.
At $t=0$ the Morse potential is switched on in the bubble, and the dynamics is followed until the bubble closes. The cut-off value $\rho^*$ of the inter-base distance $\rho$ defining closed (for $\rho<\rho^*$) and open (for $\rho>\rho^*$) states is fixed to $1.19$~nm. Output values are then calculated every 1~ns, and samples are made of about 200 runs. Error bars are standard errors.

\section{Bubble closure dynamics}

In Fig.~\ref{fig1}(a) and (b) are shown typical evolutions of the bubble size, $L(t)/L(0)$, for $\beta\kappa_{\phi}=200$ [Fig.~\ref{fig1}(a)] and 300 [Fig.~\ref{fig1}(b)]. Two other geometric quantities related to the bending and twist stored in the bubble are shown: the scalar product ${\bf \hat{n}}_{\rm i}\cdot{\bf \hat{n}}_{\rm e}$, where ${\bf\hat{n}}_{\rm i}$ and ${\bf \hat{n}}_{\rm e}$ are the tangent vectors of both dsDNA arms (see snapshot in Fig.~\ref{fig2}), and the mean twist angle per base-pair inside the bubble
\be
\Delta \phi (t) =\frac1{L(t)} \sum_{i=i_0}^{i_0+L(t)-1}\phi_i(t)
\ee
where $i_0$ is the bubble first monomer.

Two regimes can be clearly distinguished for any $L(0)$ and $\kappa_\phi$: first a \textit{zipping} regime, where $L(t)/L(0)$ decreases rapidly, until it reaches a second \textit{metastable} regime characterized by a stationnary $L(t)=\bar{L}\approx10$~bp.
In the simulations, we defined the onset of the metastable regime as the first time $t$ such that $L(t)=11$~bp. 
\begin{figure}[!ht]
\begin{center}
(a)\includegraphics[width=0.4\textwidth]{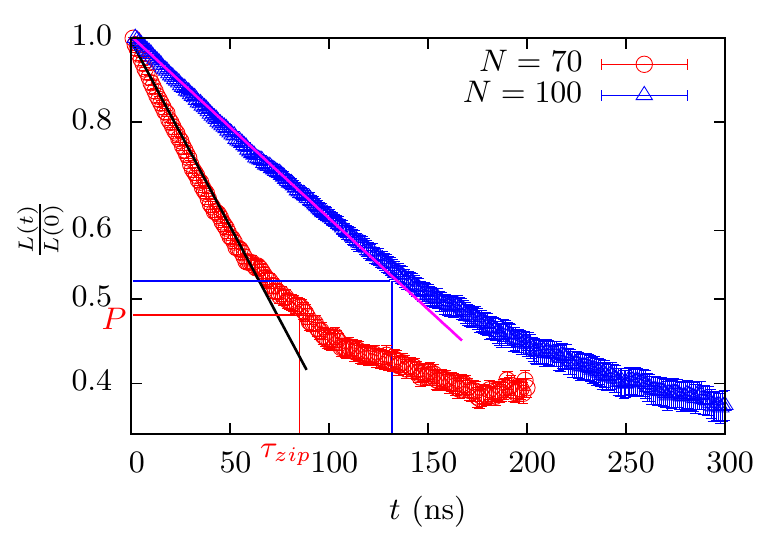}
(b)\includegraphics[width=0.4\textwidth]{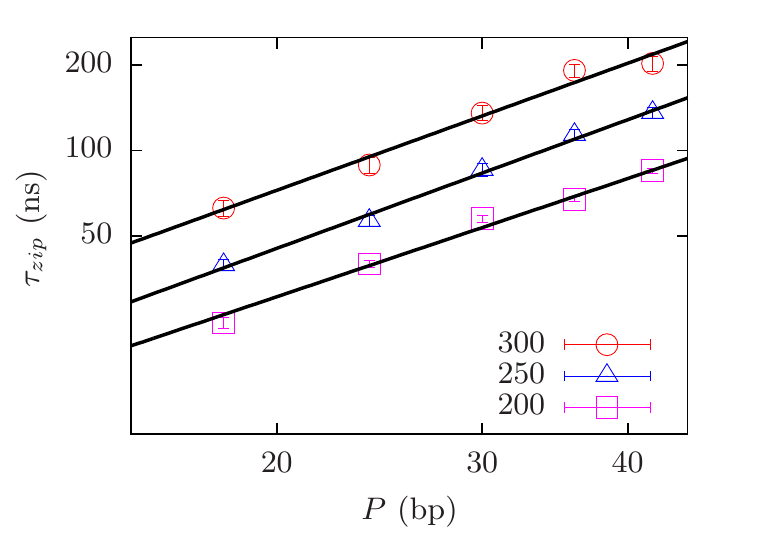}
\end{center}
\caption{\label{fig2app}(Color online)
(a) Semi-log plot of the bubble size $L(t)/L(0)$ vs. time for $\beta\kappa_{\phi}=300$, $N=70$ ($\circ$), and $N=100$ ($\triangle$). The solid lines are exponential fits. (b) Total zipping time as a function of $P$ yielding an exponent of between 1.39 (for $\beta\kappa_{\phi}=200$), 1.49 ($\beta\kappa_{\phi}=250$), and 1.51 ($\beta\kappa_{\phi}=300$).}
\end{figure}
In the fast zipping regime, the initially flexible bubble closes due to the attraction between the two strands induced by the Morse potential. 
One example of the zipping dynamics is shown in Fig.~\ref{fig2app}(a). The bubble size decays exponentially with a relaxation time on the order of $100$~ns [102 and 208~ns for $N=70$ and 100 respectively, see Fig.~\ref{fig2app}(a)]. Indeed, during zipping, the two arms rotate in opposite directions in order to increase the twist of the whole chain and thus close base-pairs with an angular velocity, $\omega\simeq \mathcal{T}/\zeta_0$ where $ \mathcal{T}\simeq 2 A\phi_{\rm eq}\simeq 4k_{\rm B}T\,$rad is the driving torque and $\zeta_0\simeq 2 \pi \eta \rho_0^2 \ell\simeq 5~k_{\rm B}T\,$ns is the rotational friction coefficient of the arms ($\ell$ is the arm initial length). We thus find $\omega\simeq 1$~rad/ns which induces zipping velocities $v\simeq p\omega/2\pi\simeq 2$~bp/ns. This rough argument yields a consistent value with the zipping velocities measured in Fig.~\ref{fig2app}(a) at short times.
By defining the zipping time, $\tau_{\rm zip}$, by $L(\tau_{\rm zip})=P\equiv \frac35[L(0)-\bar{L}]$, Fig.~\ref{fig2app}(b) shows scaling laws, $\tau_{\rm zip}\approx P^\gamma$ with $1.4\leq\gamma\leq1.5$, as already observed with the ladder model~\cite{dasanna}.
Zipping occurs whatever the initial configuration, whether the two arms are aligned or not.
\begin{figure*}[t]
\includegraphics[width=\textwidth]{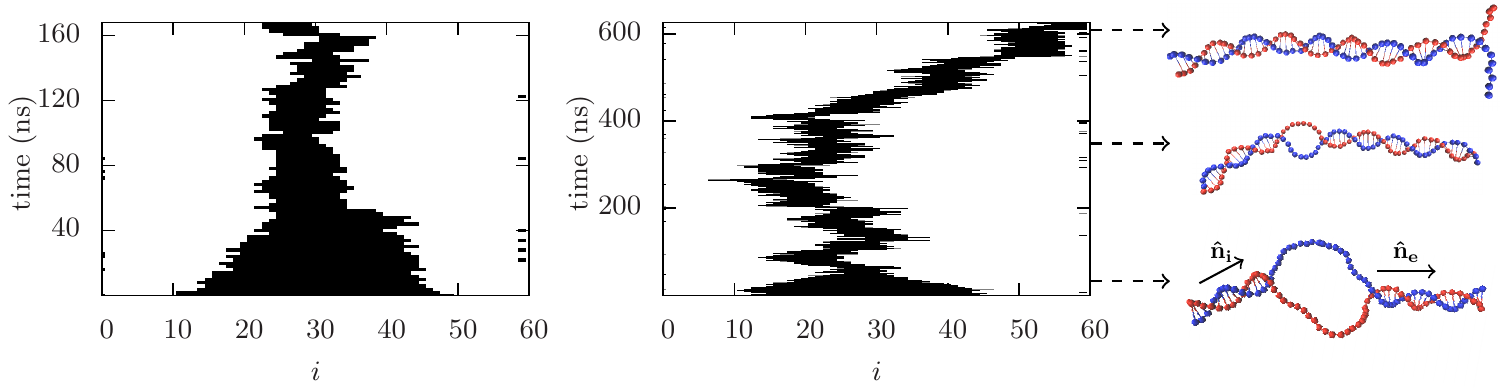}
\caption{\label{fig2} (Color online) Evolution maps (open base-pairs are in black) for $N=60$, and $\beta \kappa_\phi=200$ (left), and $\beta \kappa_\phi=300$ (middle). Three snapshots of the DNA are shown, during zipping (bottom), in the metastable state (middle), and just before BDL closure (top).}
\end{figure*}

The onset of the metastable bubble comes from the high 3D curvature of the two single strands inside the bubble, when its size reaches the ssDNA persistence length, $\bar L\approx\ell_{\rm ss}$. The two bubble single-strands are quite stiff at this scale. Either the arms are not aligned at the end of zipping and the elastic energy is both of bending and torsional nature, or they are aligned and it is only of torsional nature. The non-zero twist at the onset of the metastable state ($\Delta \phi\approx$~0.2 to 0.3~rad) is created by the fast out-of-equilibrium dynamical closure of the bubble. This is illustrated in Fig.~\ref{fig1}(c) showing the profile of the twist angle along the DNA just before (in red) and just after the onset of the metastable regime (in blue) for the simulation run shown in Fig.~\ref{fig1}(b). It clearly shows that the zipping stops as soon as the two domain walls, of approximately 5~bps width, ``collide'', increasing the twist angle value, and thus the twist energy, in the bubble centre. In brief, the zipping carries on as long as the elastic energy in the middle of the bubble is negligible. We have checked that the non-zero twist profile in the metastable state results from purely elastic properties of ssDNA (Appendix~\ref{AppB}).

Depending on the value of $\kappa_{\phi}$, the torsional contribution of the elastic energy will be an energy barrier or not. Indeed, the closure mechanism, and therefore the dwell time in the metastable state, vary with $\kappa_\phi$.
For  $\beta\kappa_{\phi}=200$ [Fig.~\ref{fig1}(a)], $\Delta\phi(t)$ increases smoothly until the bubble closes, whereas ${\bf\hat{n}}_{\rm i}\cdot{\bf\hat{n}}_{\rm e}$ increases from a negative value, to a positive one in the metastable state. The bubble closure is thus controlled by the alignment of the two stiff arms, since closure occurs as soon as ${\bf\hat{n}}_{\rm i}\cdot{\bf\hat{n}}_{\rm e}\simeq1$ (ADL closure). This behavior has already been observed in the DNA ladder model~\cite{dasanna} where no twist was present ($\kappa_\phi=0$). The final closure was controlled by the rotational diffusion of one arm with respect to the other one: the metastable dwell time scaled with the DNA mean arm length, $M$, as $\tau_{\rm met}^{\rm ADL}\sim M^\alpha$ with $2<\alpha<2.4$, and saturated at $\eta \beta \ell_{\rm ds}^3$ for $M>\ell_{\rm ds}$. 
We observe the same behavior for the helical DNA model with $\beta\kappa_{\phi}=200$, suggesting that, for this value, the twist does not play a significant role. As shown in Fig.~\ref{fig3}(b), we obtain $\tau_{\rm met}^{\rm ADL}\sim N^{2.23}$ for $\beta\kappa_{\phi}=200$ (fitted solid line).
The corresponding melting map, shown in Fig.~\ref{fig2}, illustrates that the bubble does not have sufficient time to diffuse far away from its initial position (the bubble diffusion coefficient along the DNA is $D\simeq1$~bp$^2$/ns).
\begin{figure}[!ht]
\begin{center}
\includegraphics[width=0.4\textwidth]{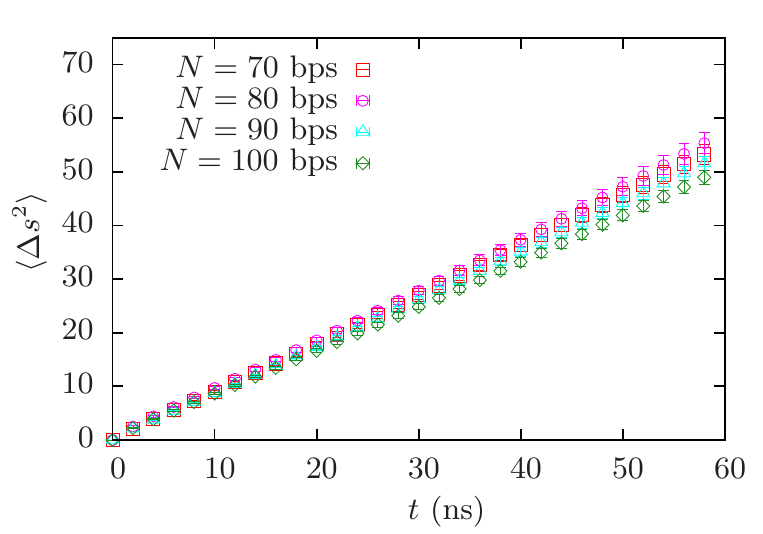}
\end{center}
\caption{\label{diffusion} (Color online) Mean-squared displacement (in units of bp$^2$) in the metastable regime of the bubble along the DNA vs. time for $\beta\kappa_{\phi}=300$ and various $N$.}
\end{figure}
For $\beta\kappa_{\phi}=300$ [Fig.~\ref{fig1}(b)] however, the arms are almost aligned during the whole metastable state (but not necessarily during zipping). Moreover, the activation barrier to continue zipping being too high (see \eq{EA} below), the fastest way to close is for the bubble to diffuse along the DNA until it reaches one end (Fig.~\ref{fig2}). 
This end opens to relax the torsion inside the bubble thus allowing a quick closure~\footnote{The second arm is ill-defined at this point, thus explaining the sudden variation of ${\bf \hat{n}}_{\rm i}\cdot{\bf \hat{n}}_{\rm e}$ just before closure in Fig~\ref{fig1}(b).}. This BDL closure time is thus controlled by the one-dimensional diffusion along the DNA.
We define precisely the arm alignment time by the condition that ${\bf \hat{n}}_{\rm i}\cdot{\bf \hat{n}}_{\rm e}=0.9$. Figure~\ref{diffusion} shows the mean-squared displacement of the bubble centre as a function of time for $\beta\kappa_\phi=300$. The bubble dynamics is purely diffusive, with a diffusion coefficient $D\simeq0.85$~bp$^2$/ns, almost independent of the DNA length.
The final closure is limited by the diffusion of the bubble towards one DNA end, leading to a dwell time in the metastable time, $\tau_{\rm met}^{\rm BDL} \approx (N/2)^2/(2D)\approx 0.15\, N^2$~ns.
Note that for $\beta\kappa_{\phi}=200$, in 25\% of the simulation runs (50 over 200) the bubble also closes using this mechanism (see Appendix~\ref{AppC}).
\begin{figure}[!ht]
\includegraphics[width=0.4\textwidth]{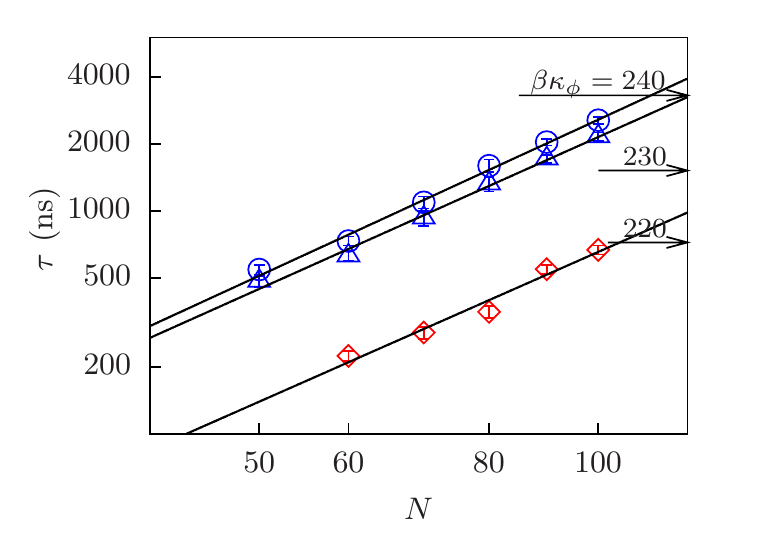}
\caption{\label{fig3} BDL dwell (\textcolor{blue}{$\triangle$}) and closure times (\textcolor{blue}{$\circ$}) ($\beta\kappa_\phi=300$), and ADL dwell times (\textcolor{red}{$\Diamond$}) ($\beta\kappa_\phi=200$) with fits (see text). Horizontal arrows are the TA dwell times for 3 values of $\beta\kappa_\phi$.}
\end{figure}
The BDL regime starts to dominate for $\beta\kappa_{\phi} >230$. Metastable dwell times, $\tau_{\rm met}^{\rm BDL}$, and closure times, defined as the first time when the bubble closes completely, $\tau_{\rm cl}^{\rm BDL}$, are plotted in Fig.~\ref{fig3} as a function of the DNA length $N$ for $\beta\kappa_{\phi}=300$. The fit yields the scaling law $\tau_{\rm met}\approx 0.06\, N^{2.3}$ which thus confirms the rough argument above (the prefactor changes due to a slightly different exponent). The total closure time follows the same scaling law $\tau^{\rm BDL}_{\rm cl}\sim N^{2.3}$. We checked that, for $\beta\kappa_{\phi}=250$, the exponent remains the same whereas the prefactor increases slightly to 0.075.
\begin{figure}[!t]
\begin{center}
(a)\includegraphics[width=0.35\textwidth]{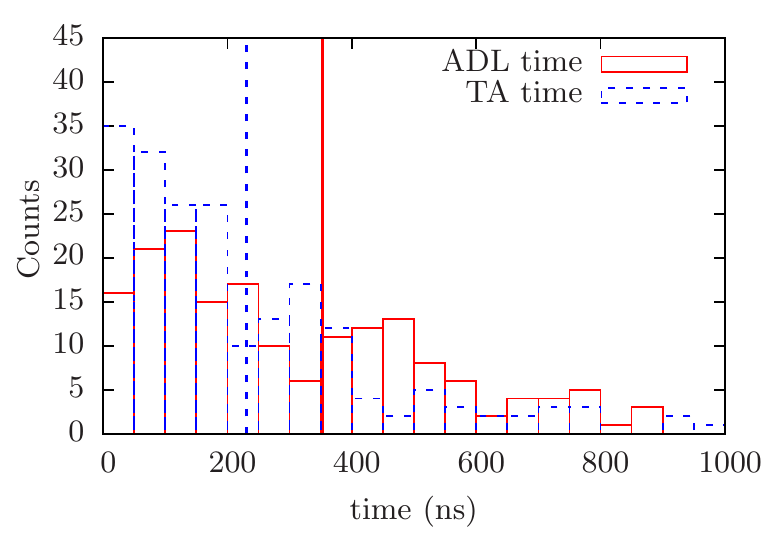}
(b)\includegraphics[width=0.4\textwidth]{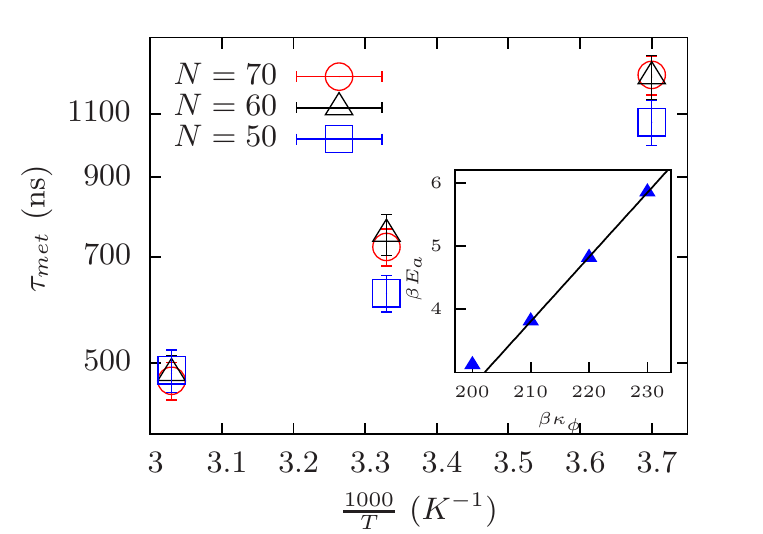}
\end{center}
\caption{\label{fig4} (Color online) (a) ADL and TA metastable dwell time distributions (free ends, $\beta\kappa_\phi=200$, $N=100$). Mean values are represented by vertical lines.  (b) Arrhenius plot of the mean dwell time for clamped ends and $\beta\kappa_{\phi}=220$. Inset: activation energy vs. $\kappa_{\phi}$ ($N=70$).}
\end{figure}

A third type of closure exists: some trajectories show a closure long after arms alignment but before the bubble reaches one end [Fig.~\ref{fig4}(a)]. This is a \textit{temperature activated} (TA) closure, associated with the crossing of an activation energy barrier. Its torsional contribution, due to non-zero twist in the bubble, is:
\be
E_{\rm tor}=\frac12\sum_{i=i_0}^{i_0+\bar L-1} (\kappa_{\phi}-\kappa_{\phi,i})\left( \phi_i - \phi_{\rm eq}\right)^2
\label{EA}
\ee 
Indeed, due to the connectivity of each DNA strand, all the base-pairs of such a small bubble can only close cooperatively.
To check this mechanism, we did simulations with clamped ends, which allowed us to avoid the BDL mechanism, and sufficiently large $\kappa_{\phi}$, to lower the ADL one. We clamped 10~bps on both DNA extremities (with Morse potential depth of $3 A/2$) to represent either long or heteropolymer DNAs with GC rich sequences on each side, as in experiments~\cite{altan-bonnet}. For $N=50$ and $\beta\kappa_\phi=300$, out of 20 realizations, 12 of them did not close before $100~\mu$s and 8 of them closed in  52~$\mu$s on average. The bubble diffuses back and forth between the clamped arms several times and eventually closes.
Figure~\ref{fig4}(b) clearly shows that the dwell time in this regime, i.e. the time actually spent by the bubble in the metastable state once the arms are aligned, follows an Arrhenius law
\be
\tau_{\rm met}^{\rm TA} = \tau_0\, \exp\left(E_{\rm a}/k_{\rm B}T\right)
\ee
where $\tau_0$ is a prefactor almost independent of $N$, and the measured activation energy is $\beta E_{\rm a}\simeq 0.10\kappa_\phi -17.6$ [inset of Fig.~\ref{fig4}(b)]. By computing $E_{\rm tor}$ using \eq{EA}, we find a comparable slope of $0.18$. For $\beta\kappa_{\phi}\lesssim200$, the activation energy starts to saturate since we enter the ADL regime.
Hence, from these simulations, it is clear that clamped DNAs, mimicking heteropolymer or long DNAs, take a long time to close, from tens to hundreds of $\mu$s. This is in quantitative agreement with the experimental results of Altan-Bonnet \textit{et al.}~\cite{altan-bonnet} where an Arrhenius law was measured with $E_{\rm a}= 7~\mathrm{kcal/mol}\approx 11~k_{\rm B}T_0$ for $N=30$. Indeed extrapolating the inset of Fig.~\ref{fig4}(b) to this value yields a torsional modulus, $\kappa_\phi=280~k_{\rm B}T_0$ ($C=3.7\times10^{-19}$~J\,nm), a value consistent with observations~\cite{smith,bouchiat,manoelJPCM}. Furthermore, the same activation energy value was measured in~\cite{altan-bonnet} for three different DNA constructs with an AT insert made of (i) a random sequence, (ii) a A track with its complementary T track, and (iii) a palindrome susceptible to form a cruciform. This is consistent with the scenario of a unique limiting step, that we show to be the formation of the 10~bps metastable bubble.
\begin{figure}[t]
\begin{center}
(a)\includegraphics[width=0.45\textwidth]{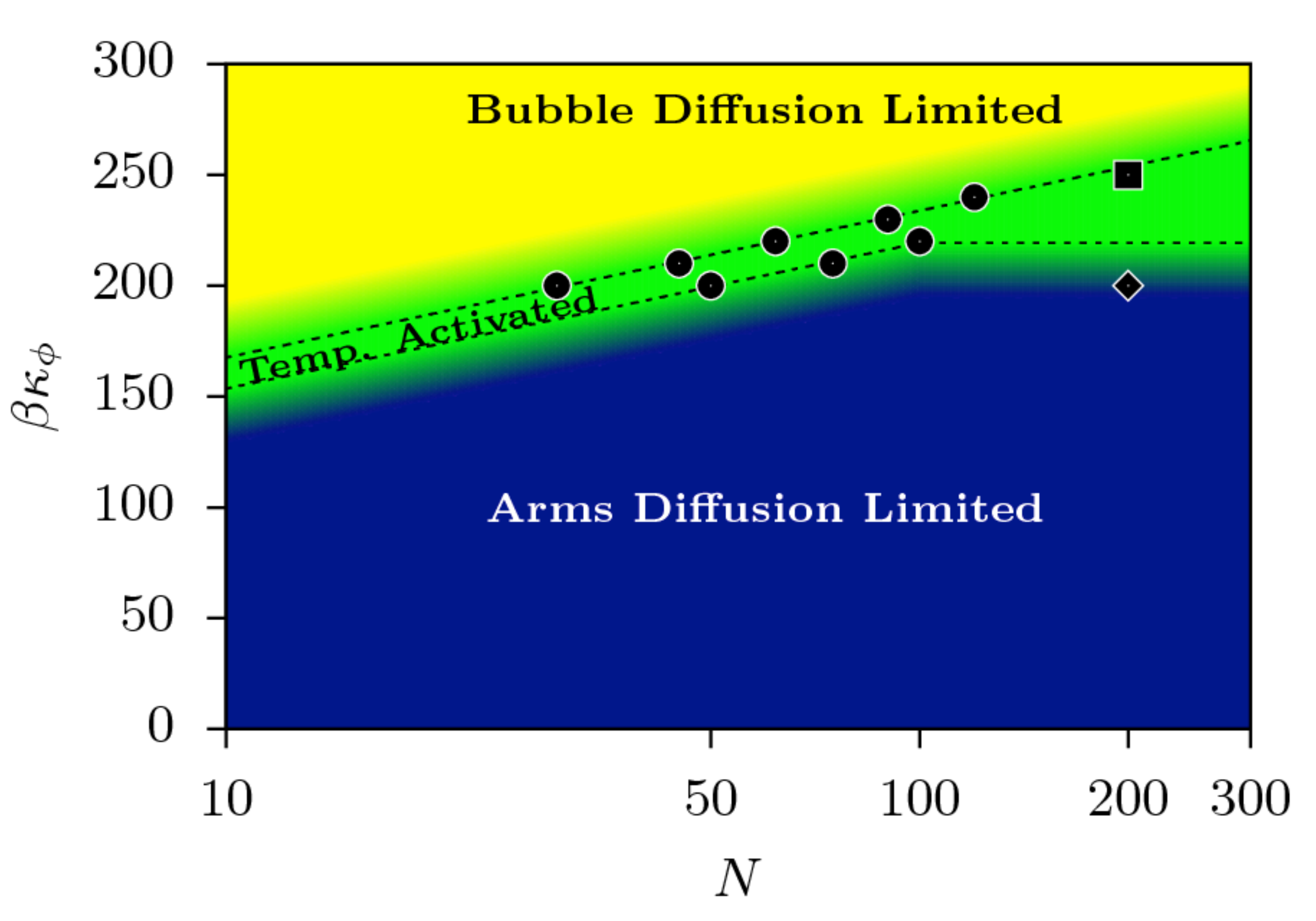}
(b)\includegraphics[width=0.45\textwidth]{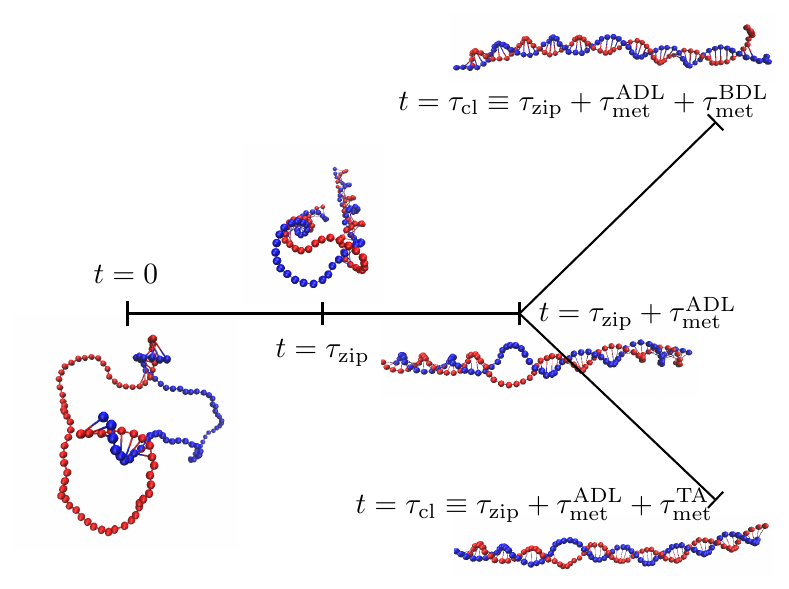}
\end{center}
\caption{\label{evol} (Color online) (a) ``Phase diagram'' of the 3 mechanisms of the bubble closure for a DNA with free ends ($\bullet$ corresponds to $\tau_{\rm met}^{\rm ADL}$ or $\tau_{\rm met}^{\rm BDL}=\tau_{\rm met}^{\rm TA}$; $\blacklozenge$ and $\blacksquare$ to simulations, see Appendix~\ref{AppC}). For clamped (e.g. GC rich) ends, the BDL region is replaced by a TA one. (b) Sketch defining the three dwell times, $\tau^{\rm ADL}_{\rm met}$, $\tau^{\rm BDL}_{\rm met}$, and $\tau^{\rm TA}_{\rm met}$ together with the zipping time, $\tau_{\rm zip}$, and the closure time, $\tau_{\rm cl}$, and corresponding snapshots.}
\end{figure}

\section{Discussion}

We performed several simulations for various $N$ and $\kappa_\phi$, and constructed a ``phase diagram'', shown in Fig.~\ref{evol}(a), representing the occurrence of the three closure regimes in the $(N,\kappa_{\phi})$ plane. The methodology used to construct this diagram is described in Appendix~\ref{AppC}. 
The definition of the various times contributing to the final closure time, $\tau_{\rm cl}$, are sketched  in Fig.~\ref{evol} with corresponding snapshots. For all the cases studied, the closure time, $\tau_{\rm zip}$, is much smaller than the dwell time in the metastable state, $\tau_{\rm met}$. In particular, as soon as the initial bubble size, $L(0)$, is larger than $\bar L$ (and equilibrated), we expect the closure time to be essentially independent of $L(0)$.

The frontiers of the different regions (ADL, BDL, and TA) should be viewed as fuzzy since the diagram is established by comparing the metastable dwell times in each regime, $\tau^{\rm ADL}_{\rm met}$, $\tau^{\rm BDL}_{\rm met}$, and $\tau^{\rm TA}_{\rm met}$, which are mean values of wide time distributions as shown in Fig.~\ref{fig4}(a). In the case of clamped (e.g. GC rich~\cite{altan-bonnet}) ends, the BDL region merges into a TA one. For realistic DNA, one can assume $\beta\kappa_\phi\gtrsim200$~\cite{smith,benhamPNAS,bouchiat}, which implies that only the BDL mechanism, for short DNAs, and the TA one, for long DNAs, might be observable. Furthermore, by extrapolating our results to very large $N$, both DNA with free ends and with clamped ends would have a bubble closure time which does not depend on $N$ any more but is controlled by the local torsion, which provides a coherent picture of bubble closure for long DNAs inside the nucleus.

A natural generalization of the model will be to consider the bubble sequence in the modeling, for instance by adjusting the parameter values in the interaction potential, $\mathcal{H}_{\rm int}$, and the torsional modulus profile, $\kappa_{\phi,i}$, with the help of the Santa Lucia's nearest-neighbor model~\cite{santalucia}.
Taking into account the single-strand torsional elasticity would slightly increase the zipping time due to elastic resistance in the bubble but would not modify the occurrence of the metastable regime. Finally, we did not consider hydrodynamic interactions in this work, and suppose the friction of the beads to be additive. The introduction of hydrodynamic interactions along one strand and between the two strands~\cite{HI} might accelerate the closure, such as it decreases the relaxation times of simple polymers. This work is in progress.
\bigskip
\bigskip

\appendix
\section{Model DNA equilibrium properties}
\label{AppA}

 This simple model captures most of the essential features of the system. The directionality is maintained by computing the sign of the determinant of $(\brho_i,\brho_{i+1},\hat n)$ and then choosing the positive sign for a right handed helix (see Fig.~\ref{fig0}).
The measured values of the geometric parameters are, after equilibration, $a_{\rm eq}=1.20a$, $\theta_{\rm eq}=0.435$~rad, and $\phi_{\rm eq}=0.52$~rad, that is slightly larger than the prescribed values $a_0=1.05a$, $\theta_0=0.41$~rad, and $\phi_0=0.62$~rad due to thermal fluctuations and non-linear potentials entering the Hamiltonian. Moreover, our model DNA is a symmetric double-helix and not a double-helix with a major and a minor grooves. The ratio contour length/axis length is equal to 1.35 in our simulations, whereas it is equal to 1.7 for a real DNA~\cite{alberts}.
\begin{figure}[ht]
\begin{center}
\includegraphics[width=.4\textwidth]{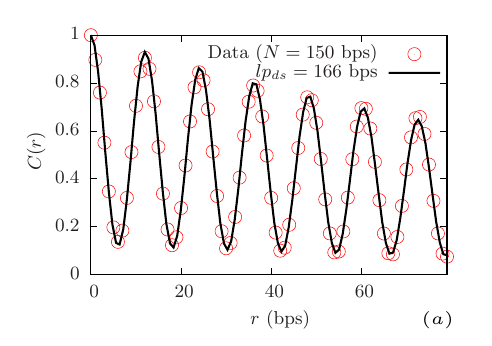}
\includegraphics[width=.4\textwidth]{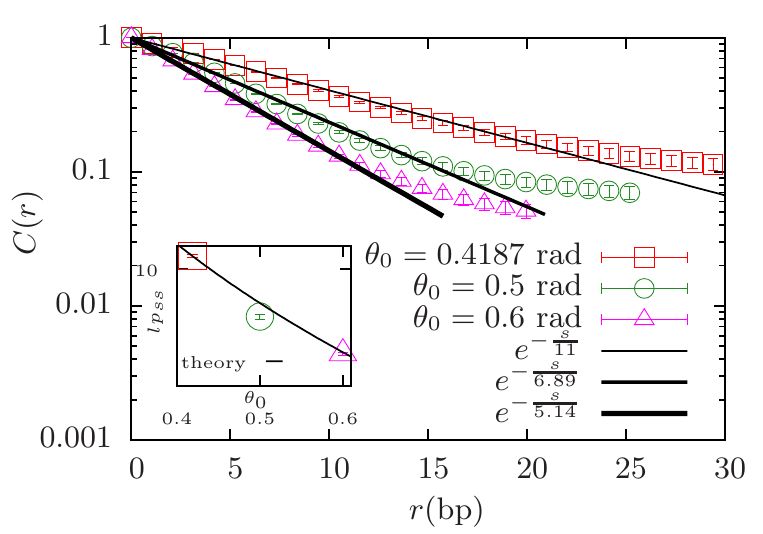}
\end{center}
\caption{\label{lp} (Color online) (a) Correlation function $C(s)$ for the simulated dsDNA ($N=150$~bps). The solid line corresponds to $C_{\rm th}(s)$ given in \eq{Cth}. (b) Semi-log plots of $C(s)$ for the simulated ssDNA, allowing the determination of $\ell_{\rm ss}$ for three different $\theta_0$ values. (Inset) Fitted value for $\ell_{\rm ss}$ plotted together with the theoretical prediction \eq{lss}.}
\end{figure}

The dsDNA persistence length, $\ell_{\rm ds}$, is computed using the method presented in Ref.~\cite{sayar2010} for $N=150$. The tangent-tangent correlation function $C(s) = \langle \hat \bt_{i+s} \cdot \hat \bt_i \rangle$ is computed for each strand, where $\hat \bt_i=\bt_i/|\bt_i|$ with $\bt_i=\br_{i+1}-\br_i$ is the unit vector connecting the two consecutive beads along a single strand. The correlation function is fitted, in Fig.~\ref{lp}(a), by the following theoretical expression (valid for a continuous helical chain)
\be
C_{\rm th}(s) = e^{-s/\ell_{\rm p}} \left[u + (1-u)\cos\left(\frac{2\pi s}p\right)\right]
\label{Cth}
\ee
where the persistence length $\ell_{\rm p}$, the coefficient $u$, and the helical pitch $p$ are fitting parameters.
The fitted values for the dsDNA persistence length and the pitch are $\ell_{\rm ds}\approx 160$~bp and $p=12$~bp for $\beta \kappa_\phi=300$, which are comparable to the actual dsDNA values ($\ell_{\rm ds}\approx 150$~bp and $p=10.4$~bp). Note that the equilibrium value of $p$ is slightly larger than the prescribed one $2\pi/\phi_0= 10$. We have checked that the dsDNA persistence length is controlled both by bending and torsional potentials as they modify the local stiffness. For $\beta \kappa_\phi=200$, we find  $\ell_{\rm ds}\approx 100$~bp. In the paper, we argue that the actual value for a real DNA is $\beta\kappa_\phi=280$, yielding $\ell_{\rm ds}\approx 150$~bp, as expected.

We also estimated the persistence length of ssDNA, $\ell_{\rm ss}$, for $N=80$. In fig.~\ref{lp}(b) is plotted the correlation function $C(s)$ for different values of $\theta_0$ in a log-linear plot. Due to the large value of the strength of the bending potential $\kappa_{\theta}$, one can assume in a good approximation that $\theta\approx \theta_0$, and  $\ell_{\rm ss}$ is purely controlled by the equilibrium bending angle $\theta_0$ (freely rotating chain model). The correlation function is thus fitted by the exponential $e^{-s/\ell_{\rm ss}}$, which yields $\ell_{\rm ss}=11$~bp, as shown in fig.~\ref{lp}(b). Moreover, we check that the ssDNA persistence length follows the law 
\be
\ell_{\rm ss}=-a/\ln(\cos\theta_0)\simeq 2a/\theta_0^2
\label{lss}
\ee
for different values of $\theta_0$ [inset of fig.~\ref{lp}(b)], as expected~\cite{grosberg}. The value, $\ell_{\rm ss}=3.7$~nm, is larger than the commonly accepted value of 1~nm. However the ssDNA persistence length is not precisely known, since it has been shown experimentally~\cite{tinland} and theoretically~\cite{Manghi2004} that it varies with the salt concentration. Values of the order of 4~nm have even been found experimentally by gel electrophoresis~\cite{tinland}. The ssDNA persistence, $\ell_{\rm ss}$, cannot be modified in our numerical model without changing the pitch value because $\theta_{\rm eq}$ is a direct function of $\phi_0$.

\section{Geometry of the metastable bubble}
\label{AppB}

We have checked that the finite value of $\Delta \phi\simeq 0.3$, or the non-zero twist profile, in the metastable bubble results from purely elastic properties of ssDNA.
We did some simulations to check the dependence of $\Delta \phi$, in the metastable state, on the ssDNA elastic parameters $\theta_0$ and $\kappa_s$. The procedure is as follows: we choose a snapshot of a metastable bubble for $N=60$ and $\beta\kappa_{\phi}=300$. Then we switch the Morse potential off inside the 10~bp bubble and slightly decrease the temperature from $T=T_0=300$~K to 0~K (15~K every $30$~ns). We observed that $\Delta \phi$ remained equal to $0.3$, thus confirming that the origin of this value is purely elastic in nature (and not entropic).

Furthermore, to find the dependence on $\Delta \phi$ with $\theta_0$, we varied $\theta_0$ from 0.3 to 0.6 (without random forces for monomers belonging to the bubble). We found the linear law $\Delta \phi= 0.53 \,\theta_0+0.06$.
This is reminiscent of the 3D bending of an elastic rod (see e.g. Ref.~\cite{landau_elasticity}) with a spontaneous curvature $\theta_0$.

Note that by decreasing the stretching modulus, $\kappa_s$, we also observed an increase of $\Delta \phi$ of 10\% for $\beta\kappa_s=40$, which might be a signature of the coupling between stretching and twisting as already mentioned in the literature~\cite{marko_stretch_twist}. 
Finally, we have checked that by slightly changing the value of $\lambda'$ in the profile $\kappa_{\phi,i}$ from 0.113~nm to 0.165~nm, we still observed the same metastable bubble size (data not shown).

The 3D deformation of the single strands in the bubble comes form the constraint at their ends. They take a fluctuating helical conformation from which we can distinguish two elastic contributions: (i) the bending is associated to the curvature of the central axis of the bubble, and (ii) the torsion is associated to the helical curvature of the strands, the central axis of the bubble remaining straight. Hence, the mean twist stored in the bubble in the metastable state results from a 3D bending of the bubble single-strands.

\section{Phase diagram construction}
\label{AppC}

\begin{figure}[t]
\begin{center}
(a)\includegraphics[width=0.4\textwidth]{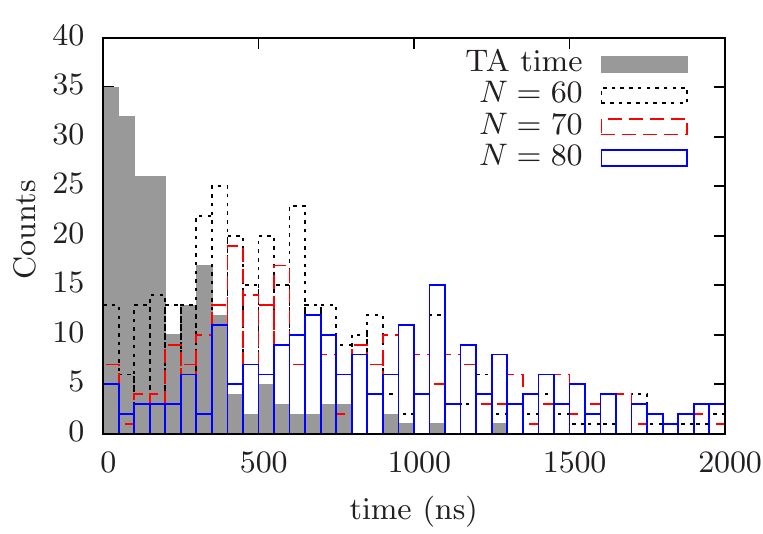}\\
(b)\includegraphics[width=0.4\textwidth]{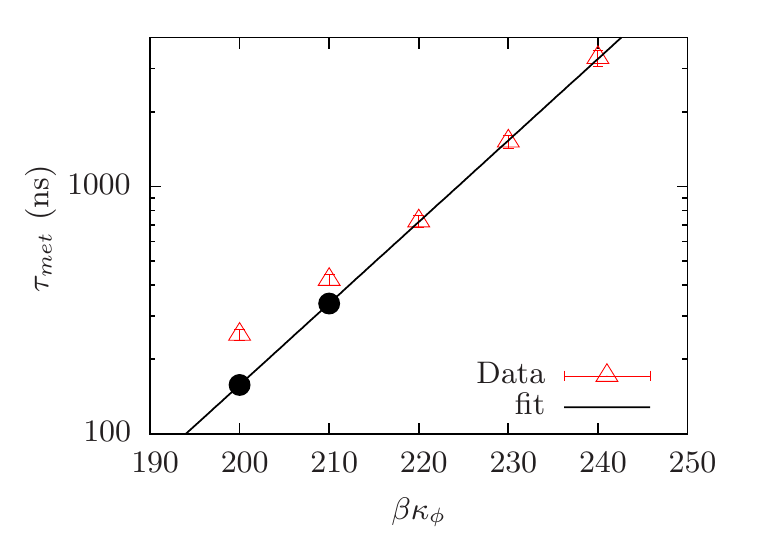}
\end{center}
\caption{\label{distribution} (Color online) (a)~BDL dwell time, $\tau^{\rm BDL}$, distributions for various $N$ and $\kappa_\phi=300$, together with the TA time, $\tau^{\rm TA}$, distribution for free ends and $\beta\kappa_\phi=200$ [same as Fig.~4(a)]. (b)~Evolution of the TA dwell time with $\kappa_\phi$ for clamped ends. The values for $\beta\kappa_\phi=200$ and 210 ($\bullet$) are extrapolated.}
\end{figure}
In Figure~\ref{distribution}(a) are shown the dwell time distributions for BDL and TA closures. The procedure to measure them is as follows: for each trajectory, the ADL times and the TA or BDL times are measured. ADL times, $\tau^{\rm ADL}$, are elapsed times between the end of zipping ($L(t)=11$) and the arms alignment (${\bf \hat n}_i\cdot{\bf \hat n}_e=0.9$), BDL times, $\tau^{\rm BDL}$, are times between alignment and closure at one DNA end, and TA times, $\tau^{\rm TA}$, are times between alignment and closure inside the DNA (see Fig.~\ref{evol}).
\begin{table*}[!t]
\centering
\begin{tabular}{| c | C{0.95cm} | C{0.95cm} | C{0.95cm} || C{0.95cm} | C{0.95cm} | C{0.95cm} || C{0.95cm} | C{0.95cm} | C{0.95cm} || C{0.95cm} | C{0.95cm} | C{0.95cm} || C{0.95cm} | C{0.95cm} | C{0.95cm} |}
\hline
\hfill $N$&\multicolumn{3}{c||}{60} & \multicolumn{3}{c||}{70} & \multicolumn{3}{c||}{80} & \multicolumn{3}{c||}{90} & \multicolumn{3}{c|}{100} \\
\cline{2-16}
$\kappa_{\phi}$& BDL & TA & ADL & BDL & TA & ADL & BDL & TA & ADL & BDL & TA & ADL & BDL & TA & ADL \\ 
\hline  
200 &35.4&49.4&15.1&28.4&40.0&31.5&20.2&49.2&30.5&12.5&33.9&50.5& 9.1&31.7& 59.1 \\ \hline
210 &56.4&35.9&7.7&44.4&35.2&20.4&34.7&43.5&21.7&28.3&44.0&27.7&20.2&45.0&34.8 \\ \hline
220 &70.0&22.7&7.2&56.8&32.9 & 10.1&54.3&37.5&8.1&42.6&43.6&13.7&34.5&38.0&27.3 \\ \hline
240 &89.6&9.2&1.0&89.6&9.8&0.5&81.2&16.2&2.5&74.8&20.7&4.3&71.1&23.9&4.9 \\ \hline
250 &95.2&4.7&0.0&90.5&7.8&1.5&90.6&6.7&2.6&90.1&6.5&3.2&87.8&9.4&2.7 \\ \hline
300 &100 & 0 & 0 & 100 &0&0&100&0&0 &100&0&0 &100&0&0 \\ 
\hline
\end{tabular}
\caption{Percentage of  the bubble  trajectories (DNA with free ends) following the closure mechanisms, ADL, BDL or TA.}
\label{table}
\end{table*}
One clearly observes the increase of the mean value and the spreading of the distribution with increasing $N$ for the BDL case. In order to construct the phase diagram, we compare the average times of these distributions. 
All the data from simulations are given in Table~\ref{table}, where $\kappa_{\phi}$ is given in $k_{\rm B}T_0$ and
$N$ in bps. For a given $\kappa_{\phi}$ and $N$, the percentage of realizations belonging to BDL, TA and ADL are given.
The percentages are computed for $\approx 200$ realizations. To distinguish between ADL and TA closure mechanisms, we first calculated from the whole metastable trajectory, $\tau^{\rm ADL}$ and $\tau^{\rm TA}$. Then, if $\tau^{\rm ADL}>\tau^{\rm TA}$, we took that trajectory to belong to ADL case and vice versa. We used the same procedure to distinguish between BDL and TA closure mechanisms.

The above data agrees with the phase diagram. We also did a few simulations for larger $N$:\\
\noindent $\bullet$~For $\beta\kappa_{\phi}=200~\&~N=200$ ($\blacklozenge$ in phase diagram): 56 realizations; 37 ADL closures, 1 BDL closure, and 18 TA closures. This point is indeed slightly below the ADL/TA frontier line in the phase diagram of the article, as expected. The mean closure time is $\tau_{\rm cl}=1.9~(\pm0.1)~\mu$s\\
\noindent $\bullet$~For $\beta\kappa_{\phi}=250~\&~N=200$ ($\blacksquare$ in phase diagram): 56 realizations; 1 ADL closure, 23 BDL closures, and 32 TA closures. This point is thus almost at the frontier BDL/TA, as it can be checked in the phase diagram. The mean closure time is $\tau_{\rm cl}=4.7~(\pm0.4)~\mu$s.

The few assumptions made in constructing the diagram are: ADL closure does not depend on $\kappa_{\phi}$; BDL times does not depend on $\kappa_{\phi}$ [as shown in Fig.~\ref{diffusion}]; and TA times are independent on $N$ (local mechanism, see Fig.~4(b) of the main article). 
Furthermore, as one needs to know the TA times for $\beta\kappa_{\phi}=200$ and 210, since these values correspond to both ADL times and TA times (as observed in simulations), we did an extrapolation as shown in Fig~\ref{distribution}(b). All the times are plotted in the same Fig.~\ref{fig3} to extract the few data points for constructing the phase diagram.

Relevant data points of the phase diagram are extracted from Fig.~\ref{fig3}. For example, the intersection between a TA horizontal line and the ADL one for $\beta\kappa_{\phi}=200$ gives $N=52$. Hence above $N=52$, the metastable bubble closes mainly through ADL. Likewise, the intersection between a TA horizontal line for $\beta\kappa_{\phi}=240$ and the BDL one for $\beta\kappa_{\phi}=300$ (we assume it is almost the same for 240) gives $N=120$, which states that below $N=120$ the bubble closes by BDL (most of the realizations) and above which it closes by TA. 

Since TA times are given by $\tau^{\rm TA}_{\rm ms}=\tau_0\,\exp(E_a(\kappa_\phi)/k_{\rm B}T)$ and assuming that ADL and BDL times do not depend on $\kappa_\phi$,  $\tau_{\rm ms}=\tau_0'\, N^\alpha$, equating both times yields the equation of the line separating TA and ADL or BDL regions in the phase diagram, $\beta\kappa_\phi= v+w \ln N$.
By fitting the 5 data points for the frontier between BDL and TA regions, one obtains $v=88$ and $w=29$. The fitted frontier line between ADL and TA (3 points) yields $v=101$ and the same value for $w$. It is important to note that the frontier for low $N$ between BDL, ADL, and TA is very fuzzy. Since for arm lengths larger than the dsDNA persistence length, $M>\ell_{\rm ds}\approx150$, the ADL time does not depend on $N$ any more~\cite{dasanna}, the frontier becomes horizontal.

\end{document}